# Do The Solitary Cooled-Down Neutron Stars Spontaneously Disintegrate?

I.V.Aničin

*Faculty of Physics, University of Belgrade, Serbia and Montenegro*

We discuss the possibility for a cooled down and otherwise stable solitary neutron star to make a spontaneous transition to its potential black hole ground state. This fundamental process would mimic a precursorless explosion in which the emitted radiations consisting predominantly of the statistical mass spectrum of high-energy neutron-rich nuclei, gamma rays and (anti) neutrinos, would carry away some mass of the star, leaving behind the core of the star in the form of a black hole.

## 1. Introduction.

We begin by noting that the stationary states of heavenly bodies of given mass are actually their metastable states in which they reside halted by the corresponding type of inner pressure on their way towards their black hole ground states. During the lifetime in this metastable state there is in principle a possibility for the body to tunnel through the pressure barrier and collapse to its ground state, while the difference between all the quantities describing the initial metastable state and the final black hole ground state would be carried away by some quantity of material ejected in the transition. Such a process would be a fundamental complex process, the amplitude of which would be described by a single multi-vertex diagram, and it would be best described as a megascopic quantum decay. Now, since in the electrically neutral bulk matter the long-range interactions capable of inducing such a transition are absent, such processes are impossible, except perhaps for a neutron star, which is the system closest to microscopic systems governed by fundamental interactions only. Here we examine the case of a neutron star in some detail.

## 2. Phenomenology of the process

We assume that a cooled-down solitary neutron star already exists, irrespective of how it might have been formed. In spite of many uncertainties, the structure of neutron stars has been modeled extensively (for recent reviews see e.g. [1-5]) and Fig.1 presents a schematic cross-section through such a star with a mass close to the presumed upper limit of about 1.5 (to perhaps 3) solar masses. The radius of the star is of the order of 10 km, and central density is at its maximum that is not supposed to exceed $5 \times 10^{15}$ g/cm$^3$. The star is highly relativistic and is, most probably, rapidly spinning.

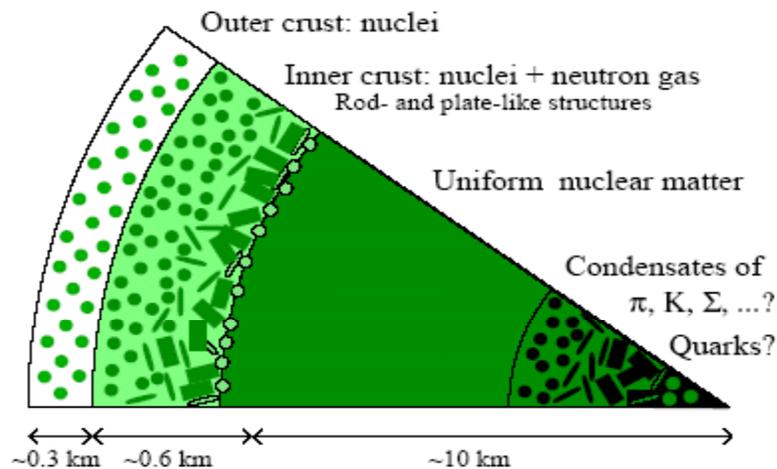

**Fig. 1**: Cross section of an ~1.4M$_\odot$ neutron star (from [5]). The interior of the star contains a nuclear liquid of mainly neutrons and ~ 10% protons at densities above nuclear matter density, increasing towards the center. Here pressures and densities may be sufficiently high that the dense cold strongly interacting matter undergoes phase transitions to quark or hyperon matter. Maximum possible density at the center is estimated at $5 \times 10^{15}$ g/cm$^3$.

It is good to recall that though gravity is the sole interaction capable of operating the transition and is the weakest of them all, it still here, due to its universality and long-range unscreened action, binds the neutrons at the surface with binding energies of the order of pion mass, what many times surpasses the nuclear binding energies, and brings the nucleons well within the action of the repulsive core of the nuclear force. As the two separated heavenly bodies, due to the nearly conservative character of gravitation (minding the weakness of gravitational radiation), cannot bind gravitationally if they are in an initial state of positive total energy unless some kinetic energy is dissipated non-gravitationally (like via the internal friction or virialization), so a neutron star cannot collapse further, over the degenerate neutron gas pressure barrier, or eventually "tunneling" through it, unless the corresponding amount of kinetic energy is dissipated non-gravitationally, while the remains of the system end-up in a gravitationally more tightly bound state of a black hole. The process may be described by an energy diagram similar to that of a radioactive decay of an unstable nucleus, which decays from an initial excited metastable state to one of the virtually infinitely many possible black hole final states, which form a quasi-continuum of possible final states, while the emitted radiations would carry away all the quantities in which the initial and final states would differ (Fig.2).

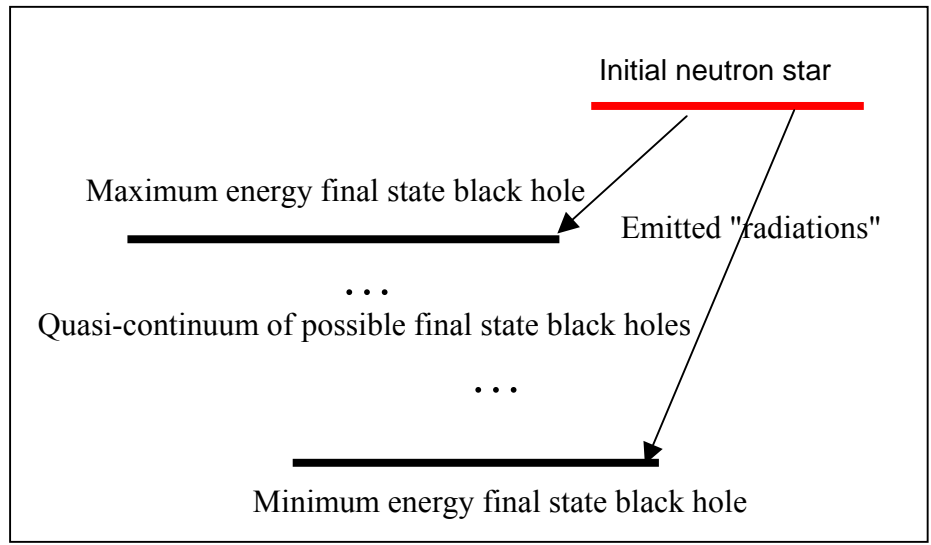

**Fig.2**. Schematic diagram of the decay of a neutron star

However, because there is no long-range force capable of coordinating the behavior of many nucleons in such a way as to yield to gravitational pressure, except gravitation itself, the realization of the process must be left to chance, or to the haphazard coincident action of all the interactions (and not to the unique unified interaction, for the energy scale here is far below that of the expected unification). We assume that within the complicated structure of the star the weak and the strong interaction would predominantly lead to virtual neutron beta decays and then to the formation of virtual nucleon clusters, and that the matter of the star would thus be boiling with virtual nuclei of different, though mostly highly neutron-rich, nuclei. Since the Pauli principle allows the protons to be space-wise in the same states and to overlap with neutrons and form virtual nuclei, when this would accidentally reach a convenient configuration throughout the body of a star, gravity might take its chance and overcome the neutron degenerate gas pressure barrier, or tunnel through it; the core would yield to gravity and collapse to its ground state black hole while the surplus nuclei would, upon getting real, be squeezed out and emitted as radiation. The nuclear binding energy liberated in the transformation of virtual nuclei into the real ones, since the nuclear force in the initial state does not bind the neutrons, and does not contribute to the binding energy of the star at all, might represent a comparatively small but maybe essential extra energy push. Since neutrons would still largely prevail the nuclei would appear highly neutron-rich, and should be, instead of the slow virialization, emitted with the corresponding kinetic energy to compensate for the increase of the gravitational total binding energy. If such a fluctuation model is realistic, then the stars that are closest to the maximum possible mass of neutron stars, which are in Fig.3 marked by an arrow, would appear most prone to the decay. The barrier between the neutron star and the black hole regions there appears the thinnest and easiest to overcome or penetrate.

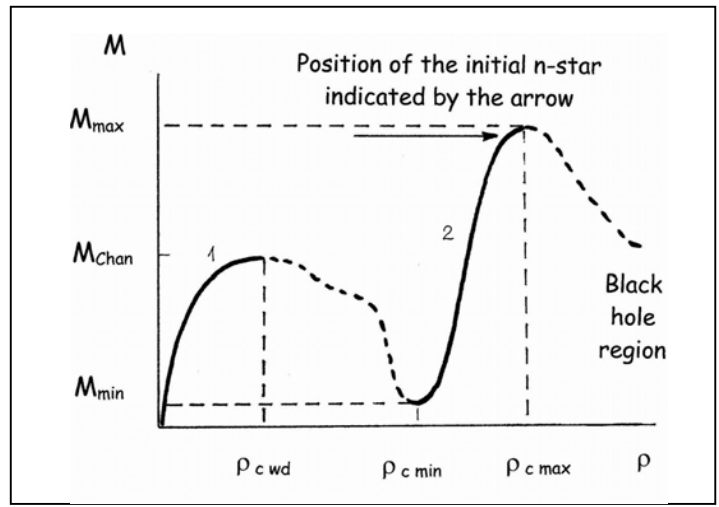

**Fig.3**. Qualitative view of the dependence of the degenerate star masses on central density (from: A. Saakian: "The Physics of Neutron Stars, Moscow 1998, in Russian). Region of white dwarfs is denoted by "1" and that of neutron stars by "2". To the right of the maximum neutron star mass-density values there is the region of black holes. The arrow points at the region where the neutron stars susceptible to "tunneling" into the black hole region are perhaps most likely to be found.

The process would look very much like a supernova explosion, with at least one important difference, that it would be absolutely precursorless and that it would have no time evolution, except for the radiations emitted in the aftermath by the unstable nuclei ejected in the decay and the effects that radiation would induce while propagating through the surrounding space. The emitted "radiations" would thus consist of the ejected neutron-rich nuclei, of the electromagnetic radiation of the broad spectrum, and of the antineutrinos rather than the neutrinos due to the virtual neutron beta decays getting real at the moment of the decay, what is opposite to supernova scenarios where fresh neutrons are needed for the production of heavy nuclei. Both the high angular momentum and the high magnetic moment of the initial state of the star would strongly influence the angular distribution of the emitted radiations. The process would deplete the population of most massive neutron stars and the count of these stars, as compared to the expectation if it is not taken into account, might reveal its presence.

**3. The amplitude of the process.**

What we can say about the amplitude of the process is of course only very general. First, we may argue on simplistic and purely probabilistic arguments, that if among the enormous number *N* of real as well as of virtual nuclear configurations, which coexist in the structure of the star, there would hopefully be a certain number *n* of those which would be favorable for the decay, the probability of the decay would be proportional to the ratio *n*/*N*. On the other hand, by analogy with microscopic quantum decays, the amplitude of the process (or the partial decay constant, or decay probability) should be bigger for the larger gravity of the star, or its mass, and for the larger phase space for the products in the final state, i.e. for the smaller number of particles and the higher energy released in the process. The available phase space per particle in the final state should thus be greatest for the smallest quantity of the material ejected when also the highest possible energy would be released and imparted to this material. The matrix elements of the transition should then also be the greatest possible for the least number of vertices would be involved. One would have to sum over the enormous number of configurations that would in the end lead to the same final state. Also, the amplitude of the transition should depend upon the similarity, or overlap, of the initial and final states – if they would greatly differ the amplitude would diminish. Now, though the concept of the radial mass distribution is for a highly relativistic object an ill-defined concept, due to the non-local character of the gravitational energy contribution to mass, we may still estimate that the high-density core of the of the n-star with its maximum density of up to 5e15 g/cm$^3$, and the radius of some ten kilometers, is not far from the gravitational radius of the final state black hole, and that the initial and final states do significantly overlap. Among the quasi-continuum of possible final states there should exist the black hole state of maximum as well as of minimum possible mass, as depicted in Fig.2. The highest energy level of the final state black hole, but lower than the initial n-star, should exist, for if it would not the star would collapse instantly to the black hole state without the need to emit anything. Also, the lowest energy state for the final state black hole different from zero should exist, for if it would not that would mean that there would be no black hole daughter left over and that the initial n-star would simply completely disintegrate into nuclei, what is again not possible. Any of the final states from the quasi-continuum between these two extremes would then be possible, and would be realized with its own partial probability $\lambda_i$ while the total probability of the decay would be $\lambda_{tot} = \Sigma_i \lambda_i$ and the mean life of the star $\tau = 1/\lambda_{tot}$.

If it exists, the process would at least partially solve the riddle of the origin of heavy, neutron-rich elements, which thus need not be synthesized at all, but would result from the fragmentation of nuclear matter formed in neutron stars, where fresh neutrons needed for the formation of heavy elements abound. Many problems that plague the synthesis of heavy elements could thus be solved (see Fig.4, and [6]).

> *Connecting Quarks with the Cosmos: 11 Science Questions for the New Century*, draft report from the National Research Council Committee on Physics of the Universe
>
> **3. How were the heavy elements from iron to uranium made?**
>
> "Scientist's understanding of the production of elements up to iron in stars and supernovae is fairly complete, but the precise origin of the heavier elements from iron to uranium remains a mystery".

**Fig.4**. The third of eleven science questions for the new century

In the end, even if the probability for the process discussed here is infinitesimally small, but still non-zero, what would make it unimportant in the large scale, in an infinite Universe it would be taking place somewhere and sometimes, and some of the violent events that we observe still might be of this nature.